\documentstyle[12pt,epsf]{article}
\begin{document}
\baselineskip 18pt
\def\baselinestretch{1.8}

\newcommand{\ba}{\begin{array}}
\newcommand{\ea}{\end{array}}
\newcommand{\bd}{\begin{displaymath}}
\newcommand{\ed}{\end{displaymath}}
\newcommand{\be}{\begin{equation}}
\newcommand{\ee}{\end{equation}}
\newcommand{\bea}{\begin{eqnarray}}
\newcommand{\eea}{\end{eqnarray}}

\def\bra{\langle}
\def\ket{\rangle}

\def\a{\alpha}
\def\b{\beta}
\def\g{\gamma}
\def\d{\delta}
\def\e{\epsilon}
\def\ve{\varepsilon}
\def\l{\lambda}
\def\m{\mu}
\def\n{\nu}
\def\G{\Gamma}
\def\D{\Delta}
\def\L{\Lambda}
\def\s{\sigma}
\def\p{\pi}

\vskip .6cm

\begin{center}
{\Large Baryogenesis through $R-$parity violation}\\
\vskip .5in
{\large Utpal Sarkar and Rathin Adhikari\footnote{
e-mail address : rathin@prl.ernet.in}}\\[.3in]

{\large Theory Group }\\
{\large Physical Research Laboratory} \\
{\large Ahmedabad - 380009, India.}

\end{center}

\vskip .75in
\begin{abstract}
\baselineskip 18pt

We  consider  generation  of  baryon  asymmetry  of the  universe
through   $R-$parity   violation   in   a   scenario   in   which
out-of-equilibrium   condition   is   satisfied   by  making  the
electroweak phase transition to be first order.  We study all the
$R-$parity  violating interaction which can generate $(B-L)$  asymmetry
which then  converts  to baryon  asymmetry  of the  universe.  We
demonstrate that CP--violating sfermion decays contribute more than
that  of the  neutralino  decays  in the  generation  of  $(B-L)$
asymmetry.

\end{abstract}

\newpage
\section{Introduction}

The  baryon  asymmetry  of  the  universe   \cite{olive}  can  be
generated at a very high energy, but in most  likelyhood  it will
be washed out at a later stage \cite{hooft,krs}.  So a great deal
of interest  started in scenarios  where the baryon  asymmetry is
generated    during    the    electroweak     phase    transition
\cite{ewbar,shapos,mstv}.  The most popular one tries to generate
the  baryon  asymmetry  in the  standard  model  or  its  minimal
extension with two higgs  doublets,  where the electroweak  phase
transition is required to be a weakly first order.  The anomalous
baryon number violation \cite{hooft} in the standard model due to
quantum effect becomes very fast at the electroweak  scale in the
presence of the  sphaleron  fields\cite{krs}.  There is provision
for  enough  $CP-$violation  in the  two  higgs  doublet  models.
Out-of-equilibrium  condition  is  satisfied  by making the phase
transition to be first order.  However, the condition  that after
the  electroweak  phase  transition  this  asymmetry  will not be
washed   out    constrains    these    models    most    severely
\cite{higgsbound}.  This  requires the higgs mass to be less than
about 80 GeV.  So if the  experimental  lower  bound on the higgs
mass is increased beyond this value, then this scenario will fail
to explain baryogenesis.

Another  interesting  scenario  has  recently  been  proposed  by
Masiero and Riotto  \cite{mas},  where they also generate  baryon
asymmetry  at around the time of  electroweak  phase  transition.
They work in the  context of  supersymmetric  model.  They  first
generte lepton number asymmetry through R-parity violating decays
of the  lightest  neutralino.  Interference  of  the  tree  level
diagram and the one loop diagram with  superparticles in the loop
gives rise to rephasing  invariant  $CP-$violation in this model.
The out-of-equilibrium  condition is satisfied by considering the
electroweak symmetry breaking phase transition to be first order.
Unlike the other class of models  \cite{ewbar,shapos,mstv},  here
the anomalous baryon number violation  converts the lepton number
asymmetry  \cite{ht,fy1} (and hence $(B-L)$  asymmetry) to baryon
asymmetry during the electroweak  phase  transition.  As a result
anomalous baryon number  violation is required to be present even
after the  electroweak  phase  transition  and hence  there is no
upper bound on the higgs mass.

In this model since lepton  number  asymmetry is generated at the
electroweak scale, the bounds \cite{nb1} on the mass of the heavy
right handed  neutrinos from  baryogenesis,  which arise from the
decay of these  particles are not valid.  The direct bound on the
masses of the  left-handed  neutrinos  and the bound on the right
handed neutrinos arising from the scattering  processes involving
the right handed neutrinos are still valid \cite{nb2}.  Otherwise
the lepton  asymmetry  generated by the decay of  neutralinos  or
sfermions would be washed out by these processes  before they are
converted to baryon asymmetry.

In this  article we point out that in  addition  to the  lightest
neutralino,  the sfermions can also  contribute to the generation
of the  lepton  asymmetry  and  hence  baryon  asymmetry  of  the
universe  in the  model of ref.  \cite{mas}.  We assume  that the
sfermions are not too heavy  compared to the mass of the lightest
neutralino.  As a  result  when  the  neutralinos  are  generated
through  the  decay  of  the  false  vacuum,  it  also   produces
sfermions, which in turn, contributes to the generation of baryon
asymmetry. We
shall not repeat the details of the model  \cite{mas}.  We shall
study all  possible  diagrams  which can generate  lepton  number
asymmetry  in the  $R-$parity  violating  models  and  hence  can
contribute to the generation of the baryon asymmetry.

In the next section we review the model in brief pointing out how
the decay of the sfermions can  contribute  to the  generation of
baryon  asymmetry.  In the following  section we describe all the
diagrams  contributing  to the generation of lepton  asymmetry in
this scenario.  The  amplitudes for these  diagrams are
then  computed and it is shown that  in  many  cases  the
contribution of the decay of the superparticles are more than
that of the
lightest  neutralino.  We then  summarize  our result in the last
section.

\section{The Model}

In this section we shall  describe only the relevant  features of
the model of ref.  \cite{mas} in brief and then point out why the
decay of the sfermions can also  contribute to the  generation of
the baryon  asymmetry  of the  universe.  The  electroweak  phase
transition  is  assumed  to be first  order.  This means  that if
$T_0$ is the  temperature  at which the  potential is flat at the
origin, and $v(T_0)$ is the vacuum  expectation  value ($vev$) of
the lightest  higgs at $T=T_0$,  then  $v(T_0)/T_0$  is non-zero.
Supersymmtry   is  broken  at  a  scale  much  larger   than  the
electroweak  symmetry  breaking  scale.  As  a  result  only  one
combination  of the higgs fields ($h$) remains light, whose $vev$
breaks the electroweak symmetry.

In this case baryogenesis is not generated before the electroweak
phase transition is over.  $(B-L)$  asymmetry is generated from a
lepton number violating (through  $R-$parity  violation) decay of
neutralinos   or  sfermions.  Due  to  anomalous   baryon  number
violation  this  $(B-L)$  asymmetry  will be  converted to baryon
asymmetry  of the  universe.  In models  where  baryon  number is
generated using anomalous  baryon number  violation, it generates
$(B+L)$  asymmetry.  Since any $(B+L)$  asymmetry  is then washed
out by the anomalous  baryon number  violation, they require that
anomalous  baryon number  violation  after the electroweak  phase
transition is too weak to wash out the generated baryon asymmetry
of the  universe.  In the  present  model  under  discussion  the
generated $(B-L)$ asymmetry is not washed out by anomalous baryon
number  violation  and hence  even  after the  electroweak  phase
transition  anomalous  electroweak baryon number violation should
be present, and there is no lower bound on $v(T_0)/T_0$.  Because
of this the  generated  baryon  asymmetry  will not be washed out
soon after the electroweak phase transition and hence there is no
lower bound on the higgs mass.

At a very high  temperature  compared  to the  electroweak  phase
transition  temperature, there is only one phase and the universe
is in the symmetric  phase.  At the critical  temperature  $T_c$,
the free energy of the $SU(2)_L  \times  U(1)_Y$  broken phase is
same  as  the  symmetry  restored  phase,  and  both  the  vacuum
co-exist.  However, at this temperature the phase transition does
not occur since the tunnelling probability through the barrier is
very small.  The phase  transition  takes place at a  temperature
$T_0 < T_c$, when the bubbles of true vacuum start  growing  very
fast and the barrier  separating the two phases nearly  vanishes.
During this time the  bubbles  collide  releasing  energy,  which
produces  particles  with a  distribution  far from  equilibrium.
This means that although the lepton number  violating  intraction
is otherwise in equilibrium at the electroweak  phase  transition
temperature ($T_0$), the  out-of-equilibrium  distribution of the
particles  allows to generate enough lepton asymmetry if there is
$CP-$violation.

Taking the  co-efficient of the quartic term in the light neutral
higgs boson $h$ to be of the order of  $\lambda_T  \sim  10^{-2}$
the bubble nucleation temperature  ($T_0$)will be about $150$ GeV
for a higgs  mass of about  100 GeV.  Because  of the  difference
between  the  false  and  the  true   vacuum   energy   densities
($\rho_{v}$),  the false  vacuum  ($\langle  h \rangle = 0$) will
decay and the bubbles  with true vacuum will expand  very fast at
temperature  $T < T_0$.  When these  bubbles  collide, the energy
releases  through  direct  particle  production  due  to  quantum
effects.  In ref \cite{mas} it was  considered  that at this stage
only the  neutralinos  will be generated  and their  distribution
will be far from  equilibrium.  However, as we shall argue, since
in  many   supersymmetric   models   the   masses  of  the  other
sfermions are comparable, all these  sfermions may also  
be produced when the bubbles  collide.  Since all these particles
have  decayed away long before the  nucleation  temperature $T_0$,  the
number density of the particles produced in this process are very
low and far from  equilibrium.  Depending  on the  mass of  these
particles,   the  number   density   will  be   suppressed.  This
suppression  is  only  logarithmic  and  hence  slightly  heavier
particles   will  also  be  produced   along  with  the  lightest
neutralinos almost in equal number.  As a result, in these models
if there are other lepton  number  violating  interactions  which
also allows enough  $CP-$violation, then they can also contribute
to the  generation  of lepton number  asymmetry.  In fact, if the
mass  of the  neutralinos  are  not  too  small  compared  to the
sfermions, then the lepton number asymmetry generated through the
decay of the sfermions can be much larger than the lepton  number
asymmetry generated by the decay of the neutralinos.

If $f_q$ fraction of particles of type $q$ is produced during a
bubble collision, then the number density of $q$ particles produced
in the collision would be,
\begin{equation}
n_q \approx \frac{f_q \rho_v \Delta }{\gamma}
\end{equation}
where, $\Delta \approx 6\sqrt{2} ({\lambda_T}/{\alpha T_0})$ is 
the size of the wall moving with a velocity $v_w$ 
and $\gamma \approx (1 - v_w^2 )^{-1/2}$. An estimate of $f_q$ is
given in ref \cite{mas} to be,
\begin{equation}
f_q \approx g_q^4 \:\:\: {\rm ln} \left(\frac{\gamma}{2 \Delta m_q}
\right) .
\end{equation}
where, $g_q$ is the Yukawa coupling constants for the higgs with the
fermions of species $q$. With this estimate 
of the number density of the particle of type $q$
produced in the collision it is possible to calculate the amount of
lepton number asymmetry generated from the decay of these particles
of species $q$. 

      From this expression one can guess that it is possible to create 
particles of mass upto ${1 \over 2} \gamma \Delta^{-1}$ when energy
is released during the collision of bubbles. In ref. \cite{mas}
the mass of  
the lightest neutralino has been taken to be about 500 GeV, 
for the choice of parameters considered. However, while 
assuming an order of magnitude of $\lambda_T$ a factor of two is not
very crucial and hence particles of masses of about 1 TeV are equally
probable. In most supersymmetric models it is assumed that several
of the sfermions will have mass less than 1 TeV. All these 
particles will then be created when energy is released during the 
collision of bubbles after the nucleation temperature $(T_0)$.

Let us assume that $CP$ is violated in the decay of these particles
$q$, and the amount of $CP$ violation is $\e_L^q$. Then the total
amount of lepton asymmetry created when these particles $q$ are created 
in collisions and then they decay is,
\begin{equation}
\frac{n_L}{s} = \frac{45}{2} \frac{\epsilon_L^q n_q}{\pi^2 g_* T_0^3}.
\end{equation}
Taking $\alpha \sim 10^{-2}$, $\gamma \sim 10^2$, $g_* \sim 10^2$ 
and $\lambda_T \sim 10^{-2}$, one obtains
\begin{equation}
\frac{n_L}{s} \approx 10^{-5} \epsilon_L^q
\end{equation}
The logarithmic
suppression factor due to the mass difference of the lightest neutralino 
and the other sfermions are almost negligible. Thus depending on the
couplings of the sfermions and the amount of $CP$ violation in their
decay, the other superparticles can generate more lepton asymmetry 
than the amount generated by the lightest neutralino. In fact, as we
shall show although the neutralino decays can generate barely enough
lepton asymmetry, the sfermion decay can generate quite large lepton
asymmetry, which makes this model more attractive. 

\section{Lepton asymmetry in decays of sfermions}

We shall  now list all the  lepton  number  violating  $R-$parity
violating  decays of the  sfermions,  which  can  interfere  with
suitable one loop diagram, which allows  $CP-$violation  and also
an  imaginary  integral.  For this  purpose we shall not  include    
processes, in which the decay products are any  superparticles or
other  heavy  particles  like  the  right  handed  neutrino.  The
decaying  particles are taken to be the sfermions, which, through
their decay to light quarks and leptons generate lepton asymmetry
if  there is  enough  $CP-$violation.  There  is  always  another
sfermion  in the  loop  to  ensure  the  absorptive  part of the
diagram to be  non-vanishing.  Lepton  number is  violated in all
these decays through $R-$parity  violation.  These leaves us with
not too many choices for the tree level and the one loop diagrams
contributing   to  the   generation   of  lepton   asymmetry.  
These diagrams are presented in figs [1-8].

We   start   with   the   $R-$parity   violating   part   of
the superpotential,

\begin{equation}
W = \lambda_{ijk} L^i L^j {\left(E^k \right)}^c +
\lambda^{\prime}_{ijk} L^i Q^j {\left( D^k \right) }^c +
\lambda^{\prime \prime}_{ijk}{\left( U^i \right) }^c{\left( D^j
\right) }^c {\left( D^k \right) }^c  
\end{equation}

\noindent
which gives all the $R-$parity violating decays of the
sfermions.  Here  $L$ and $Q$ are the lepton and quark doublet
superfields. $E^c$ is the lepton singlet superfield and $U^c$
and $D^c$ are the quark singlet superfields. $i, j, k$ are the
generation indices and $ \l_{ijk} = -\l_{jik}$ and
$\l^{\prime}_{ijk} = - \l^{\prime}_{ikj} $. In the above the
third term is a baryon  number  violating  term.  This one
cannot  generate  any baryon  asymmetry  simply  because there
are no one loop diagrams,  which can allow  $CP-$violation.
Furthermore,  for the stability of the proton, we can either
have baryon number violating  $R-$parity  breaking terms or the
lepton number violating $R-$parity breaking terms, but not both
types of terms.  So, in the present  scenario we only  consider
the lepton number violating $R-$parity breaking terms.

In  the four component Dirac notation we  can write the Yukawa
interactions of the lepton number violating R-breaking
Lagrangian generated by equation (5) as

\vbox{
\bea
{\cal L} &=&  \l_{ijk} \left[ {\tilde \n}^i_L {\bar e}^k_R e^j_L +
{\tilde e}^j_L {\bar e}^k_R \n^i_L + {\left( {\tilde e}^k_R
\right)}^* { \left( {\bar \n}^i_L \right) }^c e^j_L - \left( i
\leftrightarrow j \right) \right] \nonumber \\
&+&  \l^{\prime}_{ijk} {\mbox [  }{\tilde \n}^i_L {\bar
d}^k_R d^j_L + {\tilde d}^j_L {\bar d}^k_R \n^i_L + {\left(
{\tilde d}^k_R \right)}^* { \left( {\bar \n}^i_L \right) }^c
d^j_L  -  {\tilde e}^i_L {\bar d}^k_R u^j_L  \nonumber \\ &+& {\tilde u}^j_L
{\bar d}^k_R e^i_L  +  {\left( {\tilde d}^k_R
\right)}^* { \left( {\bar e}^i_L \right) }^c u^j_L  {\mbox ]} +
h.c. 
\eea}

\noindent
These give all the $R-$parity violating decays of the sfermions.
There are stringent bounds on different $\l_{ijk}$ and
$\l^{\prime}_{ijk}$  from low energy processes \cite{Barg}  and
very recently the product of two of such couplings has been
constrained significantly from the   neutrinoless double beta
decay \cite{babu}  and from rare leptonic decays  of the
long-lived neutral kaon, the muon and the tau as well as from
the mixing of neutral $K$ and $B$ meson \cite{Debchou}. In most
cases it is found that the upper bound on $\l^{\prime}_{ijk}$
and $\l_{ijk}$ may be of the order of $10^{-1}$ and in some
cases this bound may be of the order of $10^{-2}$   for the
sfermion mass of order 100 GeV. For higher sfermion masses
these values are even higher. Recently H1 Collaboration
\cite{HERA} has claimed that the existence of first generation
squarks is excluded for masses up  to 240 GeV for coupling values
$\l^{\prime} \geq {\sqrt {4 \pi \alpha_{em}}}$.  The upper
bound of the product of two such couplings may  vary from
$10^{-3}$ to  $10^{-4}$ except a few cases where it may be as
low as $10^{-8}$. However  in our cases in the expression of the
lepton number asymmetry the $\l$ and $\l^{\prime}$ couplings
with   various  possible combinations of the generation indices
will be involved and to make an estimate of the asymmetry we can
consider the contributions mainly coming from the $\l$ and
$\l^{\prime}$ couplings with higher values. We have considered
$\l_{ijk}$ and $\l^{\prime}_{ijk}$ to be complex in our
discussion.

Let us first consider the two body decay $ {\tilde
d}^j_L \rightarrow d^k_R {\bar \n}^i_L $ (figure 1). 
The amount of asymmetry $\e_L^q$ is defined by

\bea
\e_L^j = \sum_{ki} {\D L} {\G \left( {\tilde d}^j_L  \rightarrow
d^k_R {\bar \n}^i_L \right)  -  \G \left(  {  \left({\tilde
d}^j_L \right)}^*  \rightarrow {\bar d^k_R}  \n^i_L \right)
\over \G \left( {\tilde d}^j_L  \rightarrow {\mbox all} \right)}
\eea

\noindent
where ${\D L} $ is the  lepton number generated in the decay $
{\tilde d}^j_L  \rightarrow d^k_R {\bar \n}^i_L $. Here and in
our subsequent discussions for other decay processes also  the
magnitude of ${\D L}$ is 1.

The L-violating two body decay rate of squark for $ {\tilde
d}^j_L  \rightarrow d^k_R {\bar \n}^i_L $ is given by

\bea
\G \left(m_q, m_{\tilde q}  \right) =  {{  \left(
\lambda^{\prime}_{ijk} \right) }^2  \over 16 \pi } 
m_{\tilde{q}} { \left( 1 -
{ m_q^2 \over m_{\tilde q}^2} \right) }^2
\eea

\noindent
where $m_q$ and $m_{\tilde q}$  are the quark mass and the
squark mass   respectively. Now to get an idea of the order of
the total decay width for squarks in the R-parity violating
scenario we like to mention that in MSSM the main decay modes
are expected to be ${\tilde q}_{L,R} \rightarrow q \chi_1^0 ;
{\tilde u}_L \rightarrow d \chi_1^+ ; {\tilde d}_L \rightarrow u
\chi_1^- $ with mass of neutralinos and charginos much
lighter than that of squarks. However in our case the mass of
neutralinos and charginos are very near to the mass of squarks
and the L-violating two body decay modes we are considering has
much higher phase space in comparison to those MSSM decay modes.
Particularly when say ${\l^{\prime}}_{122} $ coupling which may
be of the order of $4 \times 10^{-1}$ is there in equation (8)
the branching ratio for L-violating decay modes may be higher
than that for those MSSM decay modes. So to estimate the value
of $\e_L$ we may consider the order of the total decay width to
be equal to the  order of the decay width for such L-violating
two body decays. Otherwise we have to include a suppression factor
given by the ratio of R--parity breaking decay rate to the decay 
rate through neutralinos.

To find $\e_L^j$  for the decay $ {\tilde d}^j_L  \rightarrow
d^k_R {\bar \n}^i_L $   we shall consider the tree level diagram
(figure 1a)  and one loop diagram (figure 1b).  In the loop
diagram  for this decay and for   other decay processes
considered by us there are MSSM type couplings at two vertices.
Unless we consider flavour violation at one of those vertices
$\e_L$ will be zero as the imaginary part of the product of the
four  couplings associated  with the four vertex in tree and
loop diagram can be made zero by suitable redefinition of the
phase associated with the  fields. Now the flavour violation is
possible in quark-squark-neutralino (or quark-squark-gluino)
interactions \cite{fcnc} as the quark and squark mass matrices
are not simultaneously diagonal.  For example, in a basis where
the  charge-1/3 quark mass matrix is diagonal, the charge -1/3
left squark mass matrix is given by
\bea
M_{L_{\tilde{d}}}^2 = 
\left( m_L^2 \, {\bf{1}} + m_{\hat{d}}^2+c_0 \, K \, m_{\hat{u}}^2 \,
K^{\dag}    
\right)
\eea

\noindent
where $m_{\hat{d}}, m_{\hat{u}}$ are the diagonal down-and
up-quark mass matrix respectively, and K is the
Kobayashi-Maskawa matrix. $m_L$ is a flavour-blind SUSY breaking
parameter that sets the scale of squark masses.  We neglect here
left-right mixing among squarks which can potentially contribute
to the off-diagonal blocks.  The term proportional to
$m_{\hat{u}}^2$  arises as a one loop contributions  induced by
up-type Yukawa coupling with charged higgsinos.  So
$m_{\tilde{d}}^2$ cannot be simultaneously diagonal with
$m_{\hat{d}}^2$  and  flavour violation occurs  in
squark-quark-neutralino  interactions.  The coefficient $c_0$
is obtained from solving the renormalization  group equations
for the  evolution of the SUSY parameters and  the value of
which is model-dependent and needs to be restricted by SUSY
contributions to various FCNC processes.

In estimating $\e_L $ for the decay  in figure 1 and in other
cases also we shall consider the flvaour violation only in one
of the two MSSM-type vertices. For such flavor violation as for
example the left-squark-quark-neutralino interaction term in the
down sector  is

\bea
{\cal L}_{q\tilde{q}\chi_i^0} &=& 
-{\sqrt{2}} \, g \, \sum_{ij} \, \left[{\tilde q}_{iL}^{\dag} \, 
{\bar{\chi}_j^0} \, {{1-\g_5} \over 2} \, q_k
\Gamma_{ik} \, \, \left\{ T_{3i}
\, N_{j2} - \tan{\theta_w} \left( T_{3i} -e_i 
\right) \, N_{j1} \right\} \, \right]   \, \nonumber \\ 
&+& \, h.c.  
\eea

\noindent
where $\Gamma_{ik}$ is the (ik)-th element of the unitary matrix
that diagonalises the upper $3 \times 3$ block of
$m_{\tilde{d}}^2$ in equation (9).  N is the neutralino mixing
matrix, and $T_{3i}$ the third component of the the isospin of
the i-th flavour. One may consider the left-right mixing among
squarks  while considering the flavor-changing
right-squark-quark-neutralino  interaction.

For a top-quark mass $m_t =170 GeV$, the third term in the
upper-left block of $m_{\tilde{d}}^2$ is important from the
viewpoint of diagonalisation, so that for a not-too-small value
of $c_0$, the elements of $\Gamma$ are close to those of K in
magnitude.  If   we parametrize  $\Gamma_{ik}$ by writing ${{\D
m_{\tilde q}}^2 \over {m_{\tilde q}}^2} \Gamma_{ik} = c K_{ik}$
where ${\D m_{\tilde q}}^2$ is the mass-square seperation
between  the two squarks  of different flavor say ${\tilde b}$
and ${\tilde s}$. The value of $c_0$ can lie anywhere between
O(0.01) to O(0.1) according to various model dependent estimates
\cite{Hage}. For higher  ${\D m_{\tilde q}}^2$  the value of
$c_0$  also can be  higher.  Thus with average squark mass in
the 200 GeV range $c$ can lie in the range $0.05 - 0.5$.  Now the
factor ${{\D m_{\tilde q}}^2 \over {m_{\tilde q}}^2}$ in
parametrizing $\G_{ik}   $  has been considered to take into
account  the GIM-like cancellations. However in our case  in
estimating $\e_L  $ we really need  not consider  this kind of
cancellations  as what matters  is the product of four couplings
as for example in figure 1(a) and figure 1(b) and as   the similar
diagrams with different flavor of quarks and squarks  may have
quite different order of values  of $\l^{\prime}  $  couplings
for which such cancellations will not be operative. In our case
we shall approximate $\G_{ik}$ as $K_{ik} $ without GIM-like
suppression.
\bea
\e_L^j   & =& \sum_{ikm} {1 \over 2 \pi} \;   
{\rm Im}\left( \l^{\prime *}_{ijk} \l^{\prime}_{imk} A_{jm} B
\right) {\left(
\sum_{ik}  {{\mid \l^{\prime}_{ijk} \mid}^2 } {
\left( 1 - { m_{d^k}^2 \over m_{\tilde {d^j}}^2} \right) }^2
\right)}^{-1} F({\tilde d}_L^j {\tilde d}_R^k) \nonumber \\
& \approx & \sum_{ikm} {1 \over 2 \pi}   \;  {\rm Im} \left(
\l^{\prime *}_{ijk} 
\l^{\prime}_{imk}  A_{jm} B  \right)  { \left( \sum_{ik} {\mid
\l^{\prime}_{ijk} \mid}^2 \right) }^{-1} \; F({\tilde d}_L^j {\tilde
d}_R^k) \nonumber
\eea
\vspace{-1.4cm}
$$
\eqno(11a)
$$
 
\noindent
where $A_{jm}$ and $B$ are given by

$$
A_{jm}     =  {\sqrt 2} g K_{jm} \left[ -{ 1 \over 2} N_{12} + {1 \over 6}
\; \tan \theta_w \; N_{11} \right] \eqno(11b) 
$$

$$
B =   \left( { - {\sqrt 2}  \over 3} \right)   \; g \tan
\theta_w   \;
{N_{12}}^*  \eqno(11c) 
$$

\noindent
In (11a), $F({\tilde d}_R^j {\tilde e}_L^l)$ comes from the absorbtive part
of the loop integral.

To estimate the order of $\e_L$ in (11a)  we  first like to mention
that it depends highly on the order of $A_{jm}$ and $B$ which comes
from the left-d-squark, quark, neutralino  coupling and the
right-d-squark, quark, neutralino coupling of the MSSM type rather
than depending on the values of $ \l^{\prime}  $ couplings which
are both in numerator and denominator of (11a).  For a wide range
of MSSM parameters for neutralino mass ranging from  about 100
to 700 GeV with $\mid \mu \mid$ about 200 to 1000  GeV and $\tan
\beta$ from 2 to 12   the product of left-d-squark, quark,
neutralino coupling with flavor violation  and right-d-squark,
quark, neutralino 
coupling  is of the order of $K_{jm} \times 10^{-7}$ to $K_{jm}
\times 10^{-2}$ and the
left-d-squark, quark, neutralino coupling is higher in general
than the similar coupling with the right-d-squark. Particularly
with neutralino mass of the order of 270 GeV and $\tan \beta =
4$ and $ \mu = -400 $ GeV  this product is about $ \; 4 K_{jm}
\times  10^{-2} $ and with neutralino mass of  the  order of 200
GeV and $ \mu = -200  $ GeV  and $  \tan \beta = 2.5$ this
product is of the order of $ K_{jm}  \times  10^{-7} $.  If we
consider the higher value of this product the order of $\e_L^j$
may be as high as of the order of $10^{-4}$ and hence ${n_L \over
s}  \sim 10^{-9}$. If one considers
all the generation indices in place of $j$ in $ e_L^j $ the
asymmetry will be even higher.  We may consider the higher
values of  $ \l^{\prime}$ allowed by the present experiments
with  its' value of  the order of $10^{-1}$.

In the reference \cite{mas} $\e_L$ is generated from the three
body decay $\chi^0_1 \rightarrow t l_i d_k^c$ which depends
highly on the $\l^{\prime}$ parameters and according to the
reference \cite{mas} if all $\l^{\prime}$ couplings are
considered of similar order then $\l^{\prime} $ are expected to
be of the order of $8 \times 10^{-3}$ to explain the baryon
asymmetry through lepton asymmetry. But the decay processes
which we are considering in this case and elsewhere the $\e_L $
depends highly on MSSM type couplings and the out of equilibrium
condition does not give bound on the MSSM couplings as the decay
width is  controlled by $\l$ or $\l^{\prime}$ couplings at the
tree level. For higher values of the product of two MSSM type
couplings one may get higher lepton asymmetry from the
L-violating two body decay modes of squarks, sneutrino or
charged leptons than that from the three body decay of
neutralino as mentioned in reference \cite{mas}.

In case of the decay $ {\tilde d}_R^k \rightarrow  \nu^i_L
d_L^j $  from the interference of the tree level and the one loop
level diagrams in figures 2a and 2b one obtains 

$$
\e^k_L \approx { \sum_{ijm}   {\rm Im} \left(  A_{mj} B  \l^{\prime }_{ijk} 
\l^{\prime * }_{imk} \right) F({\tilde d}_R^k
{\tilde d}_L^m)  \over  4 \pi  \;   \sum_{ij} 
{\mid \l^{\prime}_{ijk} \mid}^2} 
\eqno(12)
$$

\noindent
The
contribution to lepton asymmetry from this kind of interference
will be like our previous case of left-squark decay as  same
MSSM type couplings are involved.

For the decay $
{\tilde  d}_R^k  \rightarrow  e_L^i  u_L^j  $ there are two one loop
diagrams (figures 3b and 3c) which can interfere with the tree level
diagram (figure 3a). In the loop diagram  in figure 3b we shall
consider the generation mixing in the slepton-lepton-neutralino
interaction which is very similar to  the flavor changing
squark-quark-neutralino interaction considered in earlier cases.
  If the neutrinos have non-vanishing masses, the charged
slepton  mass matrix in the left sector is given by 

$$
{M_{\tilde l}}^2 = \m^2  +  M_l M_l^{\dag} + c_0^{\prime} M_{\n}
{M_{\n}}^{\dag}
\eqno(13)
$$

\noindent
where the third term comes as radiative corrections  due to the
Yukawa couplings of left sleptons with charged Higgsinos.
$c_0^{\prime}$ is  a model-dependent parameter to be specified
by the renormalization group equations. For this term  the
generation   mixing is induced in the slepton mass matrix
\cite{fcnc} and that leads to generation mixing  in the
slepton-lepton-neutralino interactions also.       If  the
neutrinos     are     Dirac particles  the matrix $m_{\n}$ is
constrained to have small elements . However see-saw type
scenarios  with large Majorana mass entail  the possibility  of
neutrino mass parameters appearing in the Yukawa couplings  to
be of the order of tau mass \cite{biswa}.  So the mixing
particularly involving third generation will be strongest in
such cases.     In this scenario the left-slepton-lepton-neutralino
interaction is   

\vbox{
\bea
{\cal L}_{l\tilde{l}\chi_i^0}  &=& 
-{\sqrt{2}} \, g \, \sum_{ij} \, \left[{\tilde l}_{iL}^{\dag} \, 
{\bar{\chi}_j^0} \, {{1-\g_5} \over 2} \, l_k
\Gamma_{ik}^{\prime} \, \, \left\{ T_{3i}
\, N_{j2} - \tan{\theta_w} \left( T_{3i} -e_i 
\right) \, N_{j1} \right\} \,   \,\right] \nonumber \\ 
  &+& \, h.c.  \nonumber
\eea
\vspace{-1.4cm}
$$
\eqno(14)
$$}

\noindent
where $\Gamma_{ik}^{\prime}$ is a function of $c_0^{\prime}$ and
the slepton mixing matrix and is constrained from  the
experimental limits on rare decays  like  $\m   \rightarrow e \g
$ and $ \tau   \rightarrow \m \g$ \cite{gabbi}. Using the bounds
on such decays \cite{PD} and  suitably  translating the limits
given in reference \cite{gabbi} it is seen that  $\m$ decay
gives the constraint on the upper limit of $ {\G_{12}^{\prime}
\; {\D m_{\tilde l}}^2 \over m_{\tilde l}^2}$  as $10^{-3}$  and
from the tau decay such constraint  on $ {\G_{23}^{\prime} \;
{\D m_{\tilde l}}^2 \over m_{\tilde l}^2}$  is $0.2 - 0.3$.

From  the interference of the diagrams in figure 3a with those
in figure 3b and figure 3c we get
\bea
\e^k_L & \approx & { \sum_{ijm} {\rm Im} \left(C_{im} B \l^{\prime
}_{ijk} \left( - \l^{\prime 
* }_{mjk} \right)  \right) F({\tilde d}_R^k
{\tilde e}_L^m) \over 2 \pi \sum_{ij} {\mid \l^{\prime}_{ijk} \mid}^2}
\nonumber \\
& + &{ \sum_{ijm} {\rm Im} \left(  B_{kl}  D  
\l^{\prime }_{ijk} \l^{\prime * }_{ijl}
\right) F({\tilde d}_R^k
{\tilde u}_L^m) \over  2 \pi \sum_{ij} {\mid \l^{\prime}_{ijk} \mid}^2}
\nonumber
\eea
\vspace{-1.4cm}
$$
\eqno(15a)
$$
\noindent
where 
$$
C_{ij}  = \G_{ij}^{\prime} \; C \eqno(15b)
$$
$$
C     =  {\sqrt 2} \;   g \left[ -{ 1 \over 2}
N_{12} - {1 \over 2} 
\tan \theta_w \; N_{11} \right] \eqno(15c) 
$$
$$
D     =  {\sqrt 2} g  \; \left[ { 1 \over 2} N_{12} + {1 \over 6}
\tan \theta_w \; N_{11} \right] \eqno(15d) 
$$
\noindent
and 
$$
B_{kl} = B \; K_{kl} \eqno(15e)
$$
\noindent
$C_{ij}$ characterizes the flavor violating
slepton-lepton-neutralino interaction in figure 3b and $B_{kl}$
characterizes flavor violating right-squark-quark-neutralino
interaction  in figure 3c and $D$  corresponds to
up-squark-quark-neutralino vertex. For flavor violation with right-squark
we  have considered similar order of suppression as in the case
of left squark.   The first term in (15a) comes from
interference of figures 3a and 3b, while the second term
comes from interference of figures 3a and 3c.  In the same
range of MSSM   parameter space discussed in the earlier case we
find the product of two MSSM type couplings $B$ and $C_{ij}$
varies from $\G_{ij}^{\prime} \times 10^{-7}$  to
$\G_{ij}^{\prime} \times  10^{-2}$  and for the product  of two
MSSM type  couplings $B_{kl}$ and $D$  it varies from $K_{kl}
\times 10^{-7}$  to $K_{kl} \times  10^{-2}$ and as the higher
values of $\G_{ij}^{\prime}$   may be somewhat higher than 0.2
the order of $ \e_L^k$ can be as high as $10^{-3}$ from (15a).

In case of the decay  $  {\tilde  u}_L^j   \rightarrow  d_R^k
{\bar e}_L^i $ there are again two one loop diagrams (figures 4b and 4c)
interferring with the tree level diagram (figure 4a) contributing to $\e_L$.
The sum of these two contributions is given by
\bea
\e^j_L & \approx & { \sum_{ikm}  
{\rm Im} \left(  D  {C_{im}}^*   \l^{\prime * }_{ijk}  \left( -
\l^{\prime  }_{mjk}  \right) \right) F({\tilde u}_L^j
{\tilde e}_L^m) \over  2 \pi \sum_{ik} {\mid \l^{\prime}_{ijk}
\mid}^2} \nonumber \\
& + & { \sum_{ikm}  {\rm Im} \left(   
\l^{\prime  * }_{ijk} D  B_{mk}   \l^{\prime
}_{ijm}  \right) F({\tilde u}_L^j
{\tilde d}_R^m) \over  2 \pi \sum_{ik} {\mid \l^{\prime}_{ijk}
\mid}^2} \nonumber  
\eea
\vspace{-1.4cm} 
$$
\eqno(16)
$$

\noindent
The order of the product of two MSSM type couplings in  equation
(16) is like earlier cases. However the higher value of this
product in the first term in (16)  may be even $\G_{im}^{\prime}
\times  10^{-1}$  for neutralino mass  of the order of 650 GeV,
$ \tan \beta = 12$ and $\mu = - 1000$ GeV  for which higher
value  of $\e_L$ from (16) may be even more than $10^{-3}$.

There is  two body sneutrino decay $ {\tilde  \n }_L^i
\rightarrow d_R^k {\bar d}_L^j     $ shown in figure 5 leading to the
contribution to $\e_L$. Now to get an idea of the order  of  the
total decay width for sneutrino we note like the case of squark
decays here also  for light neutralinos much lighter than
sneutrino in MSSM the main decay modes are expected to be $
{\tilde \n} \rightarrow \n \chi_1^0   $. However we shall
consider the mass of the lightest neutralino to be nearer to the
mass of  the sneutrino and similarly like our cases for squark
decays we shall consider the total decay width for sneutrino to
be highly dominated by the R-parity violating decay widths for
sneutrino  decaying to $d_R^k {\bar d}_L^j$ and $e_R^k {\bar
e}^j_L$.

From the interference of diagrams in figure 5a and figure 5b
one obtains 
$$
\e^i_L \approx { \sum_{jkm} {\rm Im} \left(   \l^{\prime  *
}_{ijk} {A_{mj}}^* (- C  )  \l^{\prime 
}_{imk} \right) F({\tilde \n}_L^i
{\tilde d}_L^m)  \over  2 \pi  \sum_{jk} \left(   {\mid \l_{ijk} \mid}^2 +
{\mid \l^{\prime}_{ijk} \mid}^2 \right)}
\eqno(17)
$$
\noindent
For  a wide range of parameter space mentioned in the beginning
of this section  it is found that the product of two  MSSM type
couplings  is of the order of $ K_{mj} \times 10^{-1}$. So from
this sneutrino decay one may expect the higher possible value of
$\e_L $ for a wide range of
MSSM parameter space.  The lepton asymmetry thus generated  can
be as high as , ${ n_L  \over s } \sim 10^{-8}$.

Next we shall consider L-violating decays of selectron
     like 
${\tilde e}_L^i \rightarrow d_R^k {\bar u}^j_L$.
About the total decay width here we like to mention like our
earlier cases that the main decay mode in   MSSM is expected to
be $ {\tilde l^\pm} \rightarrow l^\pm \chi_1^0 $ for light
neutralino mass.  But in our following discussion we shall
consider its' mass to be nearer to the mass of selectron  for
which one may expect that the total L-violating decay width for
selectron  decaying to      $d_R^k {\bar u}^j_L$ and  $e_R^k
{\bar \n}_L^i  $  will dominate the total
decay width for selectron.

From the interference of the diagrams  in figures 6a and 6b
one obtains 
$$
\e^i_L \approx { \sum_{jkl}  {\rm Im} \left(   \left( -  \l^{
\prime *}_{ijk} \right)   B 
C_{li}   \l^{\prime  }_{ljk} \right)  F({\tilde e}_L^i
{\tilde d}_R^k) \over  2 \pi  \sum_{jk} \left( 
{\mid \l_{ijk} \mid}^2    + {\mid \l^{\prime}_{ijk} \mid}^2
\right)  } 
\eqno(18)
$$

\noindent
The product of two
MSSM type couplings is same as in the case of the interference of
figures 3a and 3b. The order of $\e_L$ may be as high as
$10^{-3} $. 

In our discussion we have mentioned the mass of lightest
neutralino to be nearer to the mass of
squarks or charged slepton and sneutrino.  But if
it is  somewhat lighter than the squark, charged slepton or
sneutrino mass, other heavier neutralino 
 may also be lighter than squarks or charged slepton or
sneutrino. Then for various  decay processes one may get further
one  loop diagrams   replacing the lightest neutralino
 by the other heavier neutralino in
the loop diagrams and  those will   give some further
contributions to $ \e_L $. Taking into account those probable
extra diagrams and with relatively higher values of the product
of MSSM type couplings  one may get a significant  amount
of lepton asymmetry in the scenario of ref. \cite{mas}.

\section{Summary}

We studied the model proposed  by Masiero and Riotto
\cite{mas} to generate baryon asymmetry of the universe through
lepton asymmetry where the  electroweak symmetry breaking phase
transition is of first order. In contrast to their consideration
of only the three body  decay  of lightest neutralino   we have
considered various  L-violating  two body decays of
sfermions to generate lepton asymmetry   because the sfermions
may not be light and may be generated during the decay of false
vacuum. The order of lepton asymmetry  coming from these two
body decays depends highly on the choice of  various MSSM
parameters  and to some extent on the values of $\l$ and
$\l^{\prime}  $ and may easily vary from the order of $10^{-8}$
to $10^{-11} $ in the presently allowed region of MSSM parameter
space. Particularly for the decays $ {\tilde  \n }_L^i
\rightarrow d_R^k {\bar d}_L^j     $ 
one may expect quite high lepton asymmetry of the order of
$10^{-8}  $  for  a wide range of parameter space. On the other
hand the lepton asymmetry coming from neutralino decay as
mentioned in ref. \cite{mas}  depends highly on  the values of $
\l^{\prime}$ couplings and with $\l^{\prime}  \approx 8 \times
10^{-3}  $  it can be atmost of the order of $10^{-11} $. As the
lepton asymmetry  from the  decays   of squarks, sneutrino or
charged slepton mainly  depends  on the values of the product of
MSSM type couplings which are not constrained by the out of
equilibrium condition and may be quite high so even for lower values of $\l$ or
$\l^{\prime}$ of the order of $10^{-4}$ say where MSSM type    
decays may dominate the total decay width of sfermions one can still hope
for higher lepton asymmetry. Taking into account  the lepton
asymmetry  generated by sfermion decay  alongwith the neutralino
 decays  this scenario can  produce    large baryon asymmetry
of the universe.                  

\hspace*{\fill}

\noindent
{\bf  Acknowledgement}  One of us (US)  acknowledges a fellowship
from the Alexander von Humboldt  Foundation and hospitality  from
the Institut  f\"{u}r  Physik, Univ Dortmund  during his research
stay, where part of this work was done.

\newpage

\begin{figure}[htb]
\mbox{}
\vskip 1.5in\relax\noindent\hskip .5in\relax
\includegraphics{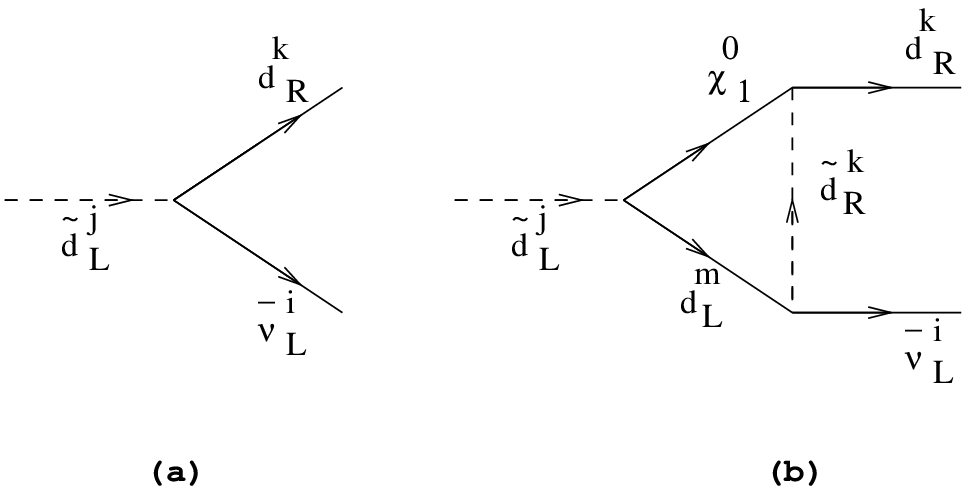} \vskip .25in
\caption{ Tree level and one loop diagram for the decay $
{\tilde d}^j_L  \rightarrow d^k_R {\bar \n}^i_L $.}
\end{figure}

\begin{figure}[htb]
\mbox{}
\vskip 1.8in\relax\noindent\hskip .5in\relax
\includegraphics{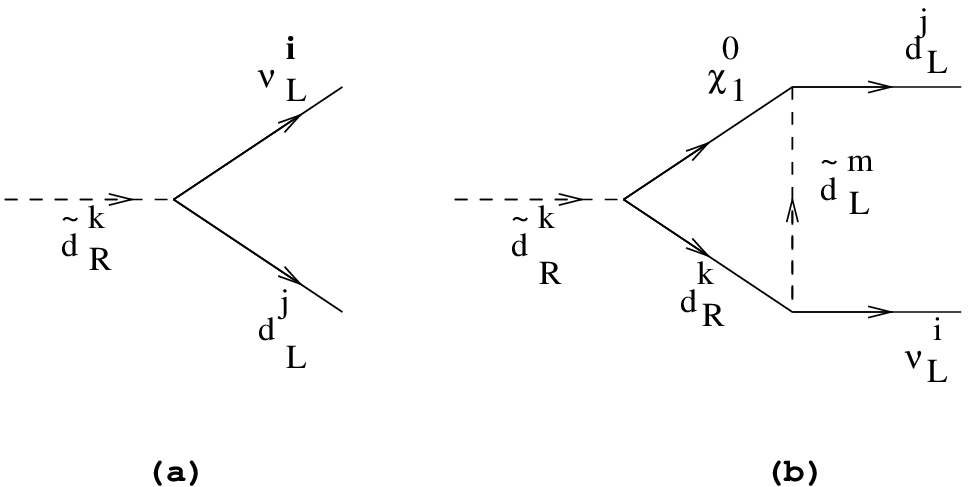} \vskip .25in
\caption{ Tree level and one loop diagram for the decay $
{\tilde d}_R^k \rightarrow  \nu^i_L
d_L^j $.}
\end{figure}

\begin{figure}[htb]
\mbox{}
\vskip 1.5in\relax\noindent\hskip -.5in\relax
\includegraphics{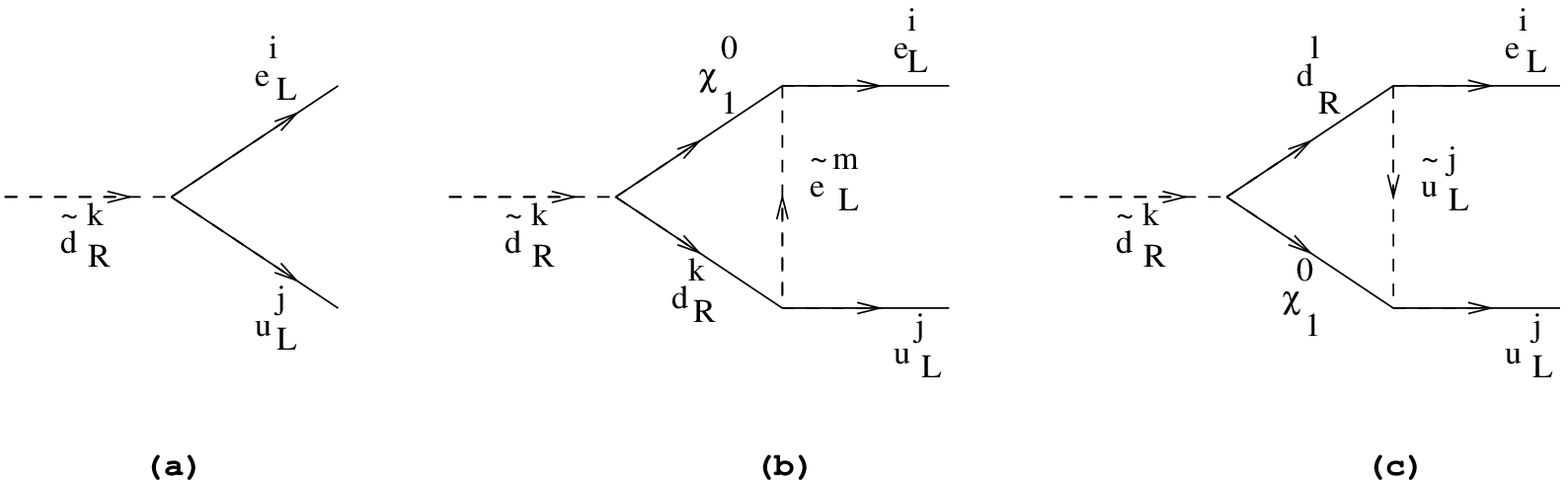} \vskip .25in
\caption{ Tree level and one loop diagram for the decay $
{\tilde  d}_R^k  \rightarrow  e_L^i  u_L^j  $.}
\end{figure}

\begin{figure}[htb]
\mbox{}
\vskip 1.8in\relax\noindent\hskip -.5in\relax
\includegraphics{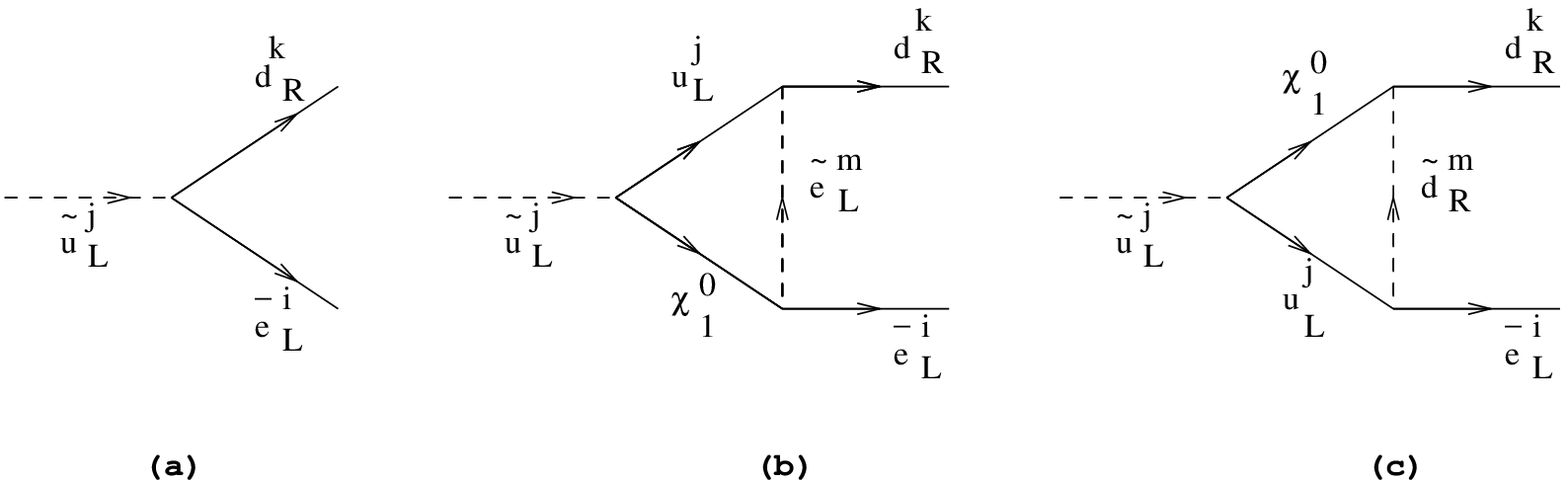} \vskip .25in
\caption{ Tree level and one loop diagram for the decay $
{\tilde  u}_L^j   \rightarrow  d_R^k
{\bar e}_L^i $.}
\end{figure}

\begin{figure}[htb]
\mbox{}
\vskip 1.5in\relax\noindent\hskip .5in\relax
\includegraphics{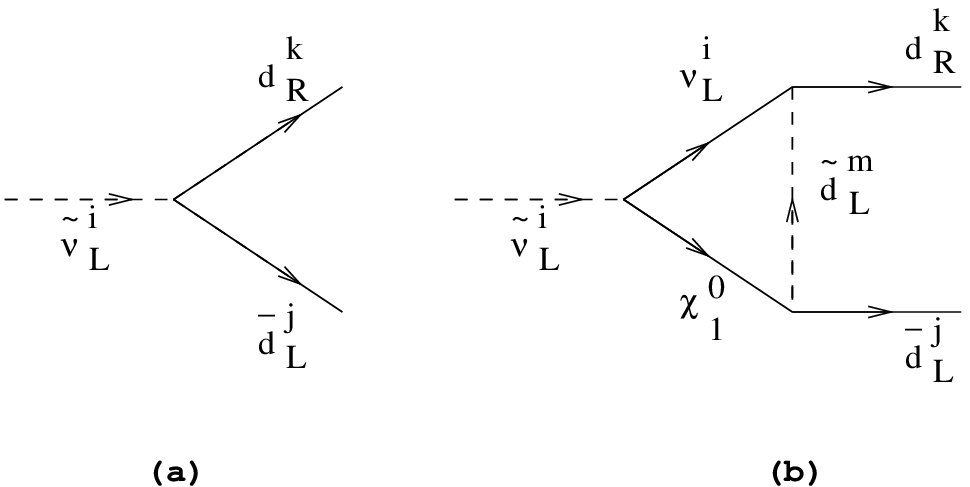} \vskip .25in
\caption{ Tree level and one loop diagram for the decay $
{\tilde  \n }_L^i
\rightarrow d_R^k {\bar d}_L^j $.}
\end{figure}

\begin{figure}[htb]
\mbox{}
\vskip 1.8in\relax\noindent\hskip .5in\relax
\includegraphics{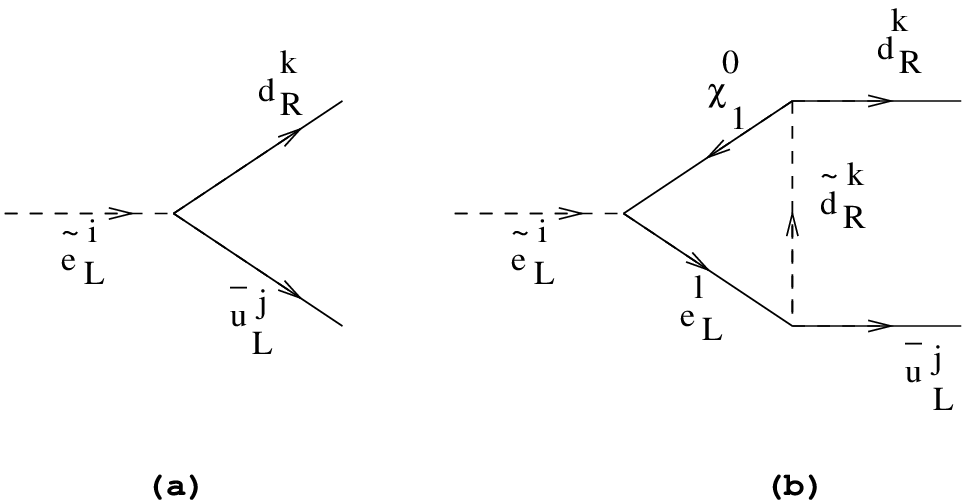} \vskip .25in
\caption{ Tree level and one loop diagram for the decay $
{\tilde e}_L^i \rightarrow
d_R^k {\bar u}^j_L$.}
\end{figure}

\end{document}